\documentclass[aps,showpacs,manuscript,12pt]{revtex4}
\usepackage{amssymb}
\usepackage{amsmath}
\usepackage{graphicx}

\begin{document}

\title{On the uniqueness of continuous inverse kinetic theory for
incompressible fluids}
\author{Massimo Tessarotto$^{1,2}$ and Marco Ellero$^{3}$}
\affiliation{$^{1}$Department of Mathematics and Informatics,
University of Trieste, Trieste, Italy \\
$^{2}$ Consortium for Magnetofluid Dynamics, Trieste, Italy\\
$^{4}$ Lehstuehl fuer Aerodynamik, Technische Universitaet
Munchen, Munchen, Germany}

\begin{abstract}
Fundamental aspects of inverse kinetic theories for incompressible
Navier-Stokes equations concern the possibility of defining
uniquely the kinetic equation underlying such models and
furthermore, the construction of a kinetic theory implying also
the energy equation. The latter condition is consistent with the
requirement that fluid fields result classical solutions of the
fluid equations. These issues appear of potential relevance both
from the mathematical viewpoint and for the physical
interpretation of the theory. In this paper we intend to prove
that the non-uniqueness feature can be resolved by imposing
suitable assumptions. These include, in particular, the
requirement that the kinetic equation be equivalent, in a suitable
sense, to a Fokker-Planck kinetic equation. Its Fokker-Planck
coefficients are proven to be uniquely determined by means of
appropriate prescriptions. In addition, as a further result, it is
proven that the inverse kinetic equation satisfies both an entropy
principle and the energy equation for the fluid fields.
\end{abstract}


\maketitle




\section{Introduction}

An aspect of fluid dynamics is represented by the class of so-called \textit{%
inverse problems}, involving the search of model kinetic theories
able to yield a prescribed set of fluid equations, with particular
reference to the continuity and Navier-Stokes (N-S) equations for
Newtonian fluids, by means of suitable velocity-moments of an
appropriate kinetic distribution function $f(\mathbf{r,v,}t)$.
Among such model theories, special relevance pertains to those
describing, self-consistently, isothermal incompressible fluids
described by the so-called incompressible Navier-Stokes equations
(INSE) in the sense that the fluid equations are satisfied for
arbitrary fluid fields and for arbitrary initial conditions for
the kinetic distribution. In particular the fluid fields are in
this case identified with $\left\{ \rho \equiv \rho
_{o},\mathbf{V,}p\right\} ,$being $\mathbf{V}$the fluid velocity
and, $\rho $and $p$respectively, the mass density and fluid
pressure, both
non-negative in the closure ($\overline{\Omega }$) of the fluid domain $%
\Omega \subseteq
\mathbb{R}
^{3},$while $\rho $is always constant (condition of
incompressibility). A desirable feature of the theory is, however,
the requirement that the relevant inverse kinetic equation are
uniquely defined \textit{(in particular, by imposing that a
particular solution is provided by local Maxwellian equilibria)}
as well as the possibility of requiring, besides INSE, also
additional fluid equations, to be satisfied by means of suitable
moment equations (\textit{extended INSE}). An example is provided
by the energy equation, i.e., the fluid equation obtained by
taking the scalar
product of the Navier-Stokes equation by the fluid velocity $\mathbf{V}$%
\begin{equation}
\frac{\partial }{\partial t}\frac{V^{2}}{2}+\mathbf{V\cdot \nabla }\frac{%
V^{2}}{2}\mathbf{+}\frac{1}{\rho _{o}}\mathbf{V\cdot \nabla
}p+\frac{1}{\rho _{o}}\mathbf{V\cdot f}-\nu \mathbf{V\cdot }\nabla
^{2}\mathbf{V}=\mathbf{0,} \label{energy equation}
\end{equation}
where $\mathbf{f}$ is the volume force density acting on the fluid
element. In fact, it is well known that the energy equation is not
satisfied by weak solutions of INSE and, as a consequence, also by
certain numerical solutions, such as those based on weak solutions
such as possibly so-called finite-volume schemes. Therefore,
imposing its validity for the inverse kinetic equation yields is a
necessary condition for the validity of
classical solutions for INSE. In a previous work \cite%
{Ellero2000,Tessarotto2004,Ellero2005}, an explicit solution to
INSE has been discovered based on a continuous inverse kinetic
theory, adopting a "Vlasov " differential kinetic equation defined
by a suitable streaming operator $L.$ Basic feature of this
kinetic equation is that, besides yielding INSE as moment
equations, it allows, as particular solutions, local kinetic
equilibria for arbitrary (but suitably smooth) fluid fluids
$\left\{
\rho _{o},\mathbf{V,}p\right\} .$ \ However, as pointed out in Refs.\cite%
{Ellero2005,Tessarotto2006}, the inverse kinetic equation defined
in this way results parameter-dependent and hence non-unique, even
in the case of local Maxwellian kinetic equilibria. This
non-uniqueness feature may result as a potentially undesirable
feature of the mathematical model, since it prevents the possible
physical interpretation of the theory (in particular, of the
mean-field force $\mathbf{F}$) and may result inconvenient from
the mathematical viewpoint since the free parameter may be chosen,
for example, arbitrarily large in magnitude. It is therefore
highly desirable to to eliminate it from the theory
\cite{Tessarotto2006}.

The purpose of this paper is to present a reformulation of the
problem which permits to cast the inverse kinetic equation in a
form which results unique, thus eliminating \ possible
parameter-dependences in the relevant streaming operator$.$
Actually, the prescription of uniqueness on the kinetic equation
is to be intended in the a suitably meaningful, i.e., to hold
under the requirement that the relevant set of fluid equations are
fulfilled identically by the fluid fields in the extended domain
$\Omega \times I$. This means that arbitrary contributions in the
kinetic equation, which vanish identically under a such an
hypothesis, can be included in the same kinetic equation.
Consistent with the previous regularity assumption, here we intend
to consider, in particular, the requirement that the inverse
kinetic equation yields also the energy equation (\ref{energy
equation}). Note that this new formulation of the inverse kinetic
theory is also relevant for comparisons both with previous
literature dealing with the determination of the probability
distribution function (PDF) for incompressible fluids
\cite{Monin1975} and with emerging theoretical approaches for the
determination of the PDF for small scale turbulence to model
homogeneous and isotropic turbulence in the inertial range. As
further development of the theory, it is shown that the streaming
operator can be suitably and uniquely modified in such a way that
the inverse kinetic equation yields the extended INSE equations,
i.e., besides the incompressible Navier-Stokes equations also the
energy equation. In particular we intend to prove that the
mean-field force $\mathbf{F}$ can be uniquely defined in such a
way that both kinetic equilibrium and moment equations yield
uniquely such equations.

\section{Non-uniqueness of the streaming operator}

We start recalling that the inverse kinetic equation which is
assumed of the form
\begin{equation}
L(\mathbf{F})f=0  \label{inverse kinetic equation-1}
\end{equation}
\cite{Ellero2005}. In particular, the streaming operator $L$ is
assumed to
be realized by a differential operator of the form $L(\mathbf{F})=\frac{%
\partial }{\partial t}+\mathbf{v\cdot }\frac{\partial }{\partial \mathbf{r}}+%
\frac{\partial }{\partial \mathbf{v}}\cdot \left\{
\mathbf{F}\right\} .$ The vector field $\mathbf{F}$
(\textit{mean-field force}) can be assumed of the
form $\mathbf{F}\equiv $ $\mathbf{F}_{0}+\mathbf{F}_{1},$ where $\mathbf{F}%
_{0}$ and $\mathbf{F}_{1},$ requiring in particular that they
depend on the minimal number of velocity moments
\cite{Tessarotto2006}, can be defined as follows

\begin{equation}
\mathbf{F}_{0}\mathbf{(x,}t;f)=\frac{1}{\rho _{o}}\left[
\mathbf{\nabla
\cdot }\underline{\underline{\mathbf{\Pi }}}-\mathbf{\nabla }p_{1}-\mathbf{f}%
\right] +\mathbf{u}\cdot \nabla \mathbf{V+}\nu \nabla
^{2}\mathbf{V,} \label{F0 non-maxwellian case}
\end{equation}
\begin{equation}
\mathbf{F}_{1}\mathbf{(x,}t;f)=\frac{1}{2}\mathbf{u}\left\{
\frac{D}{Dt}\ln
p_{1}\mathbf{+}\frac{1}{p_{1}}\mathbf{\nabla \cdot Q-}\frac{1}{p_{1}^{2}}%
\mathbf{\nabla }p\mathbf{\cdot Q}\right\} +\frac{v_{th}^{2}}{2p_{1}}\mathbf{%
\nabla }p\left\{ \frac{u^{2}}{v_{th}^{2}}-\frac{3}{2}\right\} .
\label{F1 non-Maxwellian case}
\end{equation}
Here $\mathbf{Q}$ and\textbf{\ }$\underline{\underline{\mathbf{\Pi
}}}$ are
respectively the relative kinetic energy flux and the pressure tensor $%
\mathbf{Q}=\int d^{3}v\mathbf{u}\frac{u^{2}}{3}f,$ $\underline{\underline{%
\mathbf{\Pi }}}=\int d^{3}v\mathbf{uu}f.$ As a consequence, both $\mathbf{F}%
_{0}$ and $\mathbf{F}_{1}$are functionally dependent on the
kinetic distribution function $f(\mathbf{x,}t)$. Supplemented with
suitable initial and boundary conditions and subject to suitable
smoothness assumptions for the kinetic distribution function
$f(\mathbf{x,}t)$, several important consequences follow
\cite{Ellero2005}:

\begin{itemize}
\item the fluid fields $\left\{ \rho _{o},\mathbf{V,}p\right\} $ can be
identified in the whole fluid domain $\Omega $ with suitable
velocity
moments (which are assumed to exist) of the kinetic distribution function $f(%
\mathbf{x,}t)$ [or equivalent $\widehat{f}(\mathbf{x,}t)$], of the form $%
M_{G}(r,t)=\int d^{3}vG(\mathbf{x},t)f(\mathbf{x,}t),$ where \ $G(\mathbf{x}%
,t)=1,\mathbf{v,}E\equiv \frac{1}{3}u^{2},\mathbf{v}E,$ $\mathbf{uu,}$ and $%
\mathbf{u}\equiv \mathbf{v}-\mathbf{V}(\mathbf{r,}t)$ is the
relative
velocity.\textbf{\ }Thus, we require respectively $\rho _{o}=\int d^{3}vf(%
\mathbf{x,}t),$ $\mathbf{V}(\mathbf{r,}t)=\frac{1}{\rho }\int d^{3}v\mathbf{v%
}f(\mathbf{x,}t),$ $p\mathbf{(r,}t)=p_{1}\mathbf{(r,}t)-P_{o}$ $p_{1}\mathbf{%
(r,}t)$ being the scalar kinetic pressure, i.e., $p_{1}(\mathbf{r,}t)=\int d%
\mathbf{v}\frac{u^{2}}{3}f(\mathbf{x,}t).$Requiring, $\nabla p\mathbf{(r,}%
t)=\nabla p_{1}\mathbf{(r,}t)$ and $p_{1}\mathbf{(r,}t)$ strictly
positive, it follows that $P_{o}$ is an arbitrary strictly
positive function of time,
to be defined so that the physical realizability condition $p\mathbf{(r,}%
t)\geq 0$ is satisfied everywhere in $\overline{\Omega }\times I$ ($%
I\subseteq \mathbb{R}$ being generally a finite time interval);

\item $\left\{ \rho _{o},\mathbf{V,}p\right\} $ are advanced in time by
means of the inverse kinetic equation Eq.(\ref{inverse kinetic
equation-1});

\item By appropriate choice of the mean-field force $\mathbf{F}$, the moment
equations can be proven to satisfy identically INSE, and in
particular the Poisson equation for the fluid pressure, as well
the appropriate initial and boundary conditions (see
Ref.\cite{Ellero2005});

\item The mean-field force\ $\mathbf{F}$ results by construction function
only of the velocity moments $\left\{ \rho _{o},\mathbf{V,}p_{1},\mathbf{Q,}%
\underline{\underline{\mathbf{\Pi }}}\right\} ,$ to be denoted as \emph{%
extended fluid fields}$.$

\item In particular, $L(\mathbf{F})$ can be defined in such a way to allow
that the inverse kinetic equation (\ref{inverse kinetic
equation-1}) admits,
as a particular solution, the local Maxwellian distribution $f_{M}(\mathbf{x,%
}t;\mathbf{V,}p_{1})=\frac{\rho _{0}^{5/2}}{\left( 2\pi \right) ^{\frac{3}{2}%
}p_{1}^{\frac{3}{2}}}\exp \left\{ -X^{2}\right\} .$Here, the
notation is
standard \cite{Ellero2005}, thus $X^{2}=\frac{u^{2}}{v_{th}{}^{2}},$ $%
v_{th}^{2}=2p_{1}/\rho _{o},p_{1}$ being the kinetic pressure.
\end{itemize}

Let us now prove that the inverse kinetic equation defined above (\ref%
{inverse kinetic equation-1}) is non-unique, even in the
particular case of local Maxwellian kinetic equilibria, due to the
non-uniqueness in the
definition of the mean-field force $\mathbf{F}\ $and the streaming operator $%
L(\mathbf{F})$. In fact, let us introduce the parameter-dependent
vector field $\mathbf{F}(\alpha )$
\begin{equation}
\mathbf{F}(\alpha )=\mathbf{F}+\alpha \mathbf{u}\cdot \nabla \mathbf{V-}%
\alpha \nabla \mathbf{V\cdot u}\equiv \mathbf{F}_{0}(\alpha
)+\mathbf{F}_{1} \label{F(alfa)}
\end{equation}%
where $\mathbf{F\equiv F}(\alpha =0),$ $\alpha \in \mathbb{R}$ is
arbitrary and we have denoted

\begin{eqnarray}
&&\left. \mathbf{F}_{0}(\alpha )=\mathbf{F}_{0}-\alpha \Delta \mathbf{F}%
_{0}\equiv \mathbf{F}_{0a}+\Delta _{1}\mathbf{F}_{0}(\alpha
),\right.
\notag \\
&&\left. \Delta \mathbf{F}_{0}\equiv \mathbf{u}\cdot \nabla \mathbf{V-}%
\nabla \mathbf{V\cdot u},\right.  \\
&&\left. \Delta _{1}\mathbf{F}_{0}(\alpha )\equiv (1+\alpha
)\mathbf{u}\cdot \nabla \mathbf{V-\alpha }\nabla \mathbf{V\cdot
u,}\right.   \notag
\end{eqnarray}%
where $\mathbf{F}_{0}$ and $\mathbf{F}_{1}$ given by Eqs.(\ref{F0
non-maxwellian case}),(\ref{F1 non-Maxwellian case}). Furthermore,
here we have introduced also the quantity $\Delta
_{1}\mathbf{F}_{0}(\alpha )$ to denote the parameter-dependent
part of $\mathbf{F}_{0}(\alpha )$. In fact, it is immediate to
prove the following elementary results:

a) for arbitrary $\alpha \in \mathbb{R},$ the local Maxwellian distribution $%
f_{M}$ is a particular solution of the inverse kinetic equation (\ref%
{inverse kinetic equation-1}) if and only if the incompressible
N-S equations are satisfied;

b) for arbitrary $\alpha $ in $\mathbb{R}$, the moment equations
stemming from the kinetic equation (\ref{inverse kinetic
equation-1}) coincide with the incompressible N-S equations;

c) the parameter $\alpha $ results manifestly functionally
independent of the kinetic distribution function
$f(\mathbf{x},t).$

The obvious consequence is that the functional form of the vector field $%
\mathbf{F}_{0},$ and consequently $\mathbf{F,}$ which
characterizes the inverse kinetic equation (\ref{inverse kinetic
equation-1}) is not unique. The non-uniqueness in the contribution
$\mathbf{F}_{0}(\alpha )$ is carried by the term $\alpha \Delta
\mathbf{F}_{0}$ which does not vanish even if the fluid fields are
required to satisfy identically INSE in the set $\Omega
\times I.$ We intend to show in the sequel that the value of the parameter $%
\alpha $ can actually be uniquely defined by a suitable
prescription on the streaming operator and the related mean-field
force.

\section{A unique representation}

To resolve the non-uniqueness feature of the functional form of
the streaming operator $L$, due to this parameter dependence, let
us now
consider again the inverse kinetic equation (\ref{inverse kinetic equation-1}%
). We intend to prove that the mean-field force $\mathbf{F,}$ and
in particular the vector field $\mathbf{F}_{0}(\alpha )$, can be
given an unique representation in terms of a suitable set of fluid
fields $\left\{
\rho _{o},\mathbf{V,}p_{1},\mathbf{Q,}\underline{\underline{\mathbf{\Pi }}}%
\right\} $ defined above by introducing a symmetrization condition
on the mean field force $\mathbf{F}_{0}(\alpha ).$To reach this
conclusion it is actually sufficient to impose that the kinetic
energy flux equation results parameter-independent and suitably
defined. Thus, let us consider the moment
equation which corresponds the kinetic energy flux density $G(\mathbf{x},t)=%
\mathbf{v}\frac{u^{2}}{3}$. Requiring that $f(\mathbf{x,}t)$ is an
arbitrary particular solution of the inverse kinetic equation (not
necessarily
Maxwellian) for which the corresponding moment $\mathbf{q}=\int d^{3}v%
\mathbf{v}\frac{u^{2}}{3}f$ (kinetic energy flux vector) does not
vanish
identically, the related moment equation takes the form%
\begin{eqnarray}
&&\left. \frac{\partial }{\partial t}\int
d\mathbf{v}G(\mathbf{x,}t)f+\nabla
\cdot \int d\mathbf{vv}G(\mathbf{x,}t)f-\right.   \notag \\
&&\left. -\int d\mathbf{v}\left[ \mathbf{F}_{0a}+\Delta _{1}\mathbf{F}%
_{0}(\alpha )+\mathbf{F}_{1}\right] \cdot \frac{\partial G(\mathbf{x,}t)}{%
\partial \mathbf{v}}f-\right.  \\
&&\left. -\int d\mathbf{v}f\left[ \frac{\partial }{\partial t}G(\mathbf{x,}%
t)+\mathbf{v\cdot \nabla }G(\mathbf{x,}t)\right] =0.\right.
\notag
\end{eqnarray}%
Introducing the velocity moments $p_{2}=\int d\mathbf{v}\frac{u^{4}}{3}f,$ $%
\underline{\underline{\mathbf{P}}}=\int d\mathbf{vuu}\frac{u^{2}}{3}f$ and $%
\underline{\underline{\underline{\mathbf{T}}}}=\int
d\mathbf{vuuu}f,$ the kinetic energy flux equation contains
contributions which depend linearly on the undetermined parameter
$\alpha .$ The contribution to the rate-of-change of $\mathbf{q}$
produced by $\Delta _{1}\mathbf{F}_{0}(\alpha ),$ which results
proportional both to the velocity gradient $\nabla \mathbf{V}$ and
the relative kinetic energy flux $\mathbf{Q}$, reads
\begin{equation}
\text{ \ \ \ }\mathbf{M}_{\alpha }(f)\equiv -(1+\alpha )\mathbf{Q\cdot }%
\nabla \mathbf{V}+\mathbf{\alpha \nabla \mathbf{V\cdot Q}}.
\label{moment M_alfa}
\end{equation}%
In order to eliminate the indeterminacy of $\alpha $, since
$\alpha $ cannot depend on the kinetic distribution function $f,$
a possible choice is provided by the assumption that
$\mathbf{M}_{\alpha }(f)$ takes the symmetrized form
\begin{equation}
\mathbf{M}_{\alpha }(f)=\mathbf{\mathbf{-}}\frac{1}{2}\mathbf{\nabla \mathbf{%
V\cdot Q+}}\frac{1}{2}\mathbf{Q\cdot \nabla V,}
\label{symmetruzation condition}
\end{equation}%
which manifestly implies $\alpha =1/2.$ Notice that the
symmetrization condition can also be viewed as a constitutive
equation for the rate-of-change of the kinetic energy flux vector.
In this sense, it is analogous to similar symmetrized constitutive
equations adopted in customary approaches to extended
thermodynamics \cite{muller1998}. \ On the other
hand, Eq.(\ref{symmetruzation condition}) implies $\mathbf{M}_{\alpha }(f)=%
\frac{1}{2}\mathbf{\mathbf{Q\times }\xi ,}$ $\mathbf{\xi }=\nabla
\times \mathbf{V}$ being the vorticity field. Thus,
$\mathbf{M}_{\alpha }(f)$ can
also be interpreted as the rate-of-change of the kinetic energy flux vector $%
\mathbf{Q}$ produced by vorticity field $\mathbf{\xi }$. From Eq.(\ref%
{symmetruzation condition}) it follows that $\mathbf{F}_{0}(\alpha
)$ reads
\begin{equation}
\mathbf{F}_{0}(\alpha =\frac{1}{2})=\frac{1}{\rho _{o}}\left[
\mathbf{\nabla
\cdot }\underline{\underline{\mathbf{\Pi }}}-\mathbf{\nabla }p_{1}-\mathbf{f}%
\right] +\frac{1}{2}\left( \mathbf{u}\cdot \nabla \mathbf{V+}\nabla \mathbf{%
V\cdot u}\right) +\nu \nabla ^{2}\mathbf{V}.  \label{F0}
\end{equation}%
Hence, the functional form of the streaming operator results
uniquely determined. As a result of the previous considerations,
it is possible to establish the following uniqueness theorem:\\

\textbf{THEOREM 1 -- Uniqueness of the Vlasov streaming operator }$L(\mathbf{%
F})$

\emph{Let us assume that:}

\emph{1) the fluid fields} $\left\{ \rho ,\mathbf{V,}p\right\} $
\emph{and volume force density }$\mathbf{f(r,V},t)$ \emph{belong
respectively to the
functional settings }$\left\{ \mathbf{V}(\mathbf{\mathbf{r,}}t),p(\mathbf{r,}%
t)\in C^{(0)}(\overline{\Omega }\times I),\mathbf{V}(\mathbf{\mathbf{r,}}%
t),p(\mathbf{r,}t)\in C^{(2,1)}(\Omega \times I)\right\} $\emph{\ and }$%
\left\{ \mathbf{\mathbf{f}}(\mathbf{\mathbf{r,v}},t\mathbb{)}\in C^{(0)}(%
\overline{\Omega }\times I),\mathbf{f}(\mathbf{r,}t)\in
C^{(1,0)}(\Omega \times I)\right\} ;$\emph{\ }

\emph{2) the operator }$L(\mathbf{F}),$ \emph{defining the inverse
kinetic equation (\ref{inverse kinetic equation-1})}, \emph{has
the form of the Vlasov streaming operator }$L$\emph{;}

\emph{3) the solution, }$f(\mathbf{x},t),$ \emph{of the inverse
kinetic equation (\ref{inverse kinetic equation-1}) exists,
results suitably smooth in }$\Gamma \times I$ \emph{and} \emph{its
velocity moments }$\left\{ \rho
_{o},\mathbf{V,}p_{1},\mathbf{Q},\underline{\underline{\mathbf{\Pi }}}%
\right\} $ \emph{define the fluid fields }$\left\{ \rho _{o},\mathbf{V,}%
p\right\} $\emph{\ which are classical solutions of INSE, together
with Dirichlet boundary conditions and initial
conditions}$.$\emph{\ In addition, the inverse kinetic equation
admits, as particular solution, the local Maxwellian distribution
}$f_{M}$\emph{;}

\emph{4) the mean-field force }$\mathbf{F}(\alpha )$\ \emph{is a
function
only of the extended fluid fields } $\left\{ \rho _{o},\mathbf{V,}p_{1},%
\mathbf{Q},\underline{\underline{\mathbf{\Pi }}}\right\} ,$
\emph{while the
parameter }$\alpha $\emph{\ does not depend functionally on} $f(\mathbf{x}%
,t);$

\emph{5) the vector field }$\Delta _{1}F_{0}(\alpha )$\emph{\
satisfies the the symmetry condition (\ref{symmetruzation
condition})}$.$

\emph{Then it follows that the mean-field force
}$\mathbf{F}$\emph{\ in the inverse kinetic equation (\ref{inverse
kinetic equation-1}) is uniquely defined in terms of
}$\mathbf{F}=\mathbf{F}_{0}+\mathbf{F}_{1},$\emph{where the vector
fields }$\mathbf{F}_{0}$\emph{\ and }$\mathbf{F}_{1}$\emph{\ are
given by Eqs. (\ref{F0}) and (\ref{F1 non-Maxwellian case}).}

\emph{PROOF}

Let us consider first the case in which the distribution function $f(\mathbf{%
x},t)$ coincides with the local Maxwellian distribution $f_{M}$.
In this case by definition the moments
$\mathbf{Q},\underline{\underline{\mathbf{\Pi
}}}$ vanish identically while, by construction the mean mean-field force%
\emph{\ }is given by $\mathbf{F}(\alpha )$ [see
Eq.(\ref{F(alfa)})], $\alpha \in
\mathbb{R}
$ being an arbitrary parameter. Let us now assume that
$f(\mathbf{x},t)$ is
non-Maxwellian and that its moment $\mathbf{M}_{\alpha }(f)$ defined by Eq.(%
\ref{moment M_alfa}) is non-vanishing. In this case the uniqueness of $%
\mathbf{F}$ follows from assumptions 4 and 5. In particular the parameter $%
\alpha $ is uniquely determined by the symmetry condition (\ref%
{symmetruzation condition}) in the moment $\mathbf{M}_{\alpha
}(f)$. Since by assumption $\alpha $ is independent of
$f(\mathbf{x},t)$ the result applies to arbitrary distribution
functions (including the Maxwellian case).
Let us now introduce the vector field $\mathbf{F}^{\prime }\mathbf{=F+}$ $%
\Delta \mathbf{F,}$ where the vector field $\Delta \mathbf{F}$ is
assumed to depend functionally on $f(\mathbf{x},t)$ and defined in
such a way that:

A) the kinetic equation $L(\mathbf{F}^{\prime })f(\mathbf{x},t)=0$
yields an inverse kinetic theory for INSE, satisfying hypotheses
1-5 of the present theorem, and in particular it produces the same
moment equation of the
inverse kinetic equation (\ref{inverse kinetic equation-1}) for $G(\mathbf{x,%
}t)=1,\mathbf{v,}E\equiv \frac{1}{3}u^{2}$;

B) there results identically $\Delta \mathbf{F}(f_{M})\equiv 0,$ i.e., $%
\Delta \mathbf{F}$ vanishes identically in the case of a local
Maxwellian distribution $f_{M}.$

Let us prove that necessarily $\Delta \mathbf{F}(f)\equiv 0$ also
for arbitrary non-Maxwellian distributions $f$ which are solutions
of the inverse kinetic equation. First we notice that from A and
B, due to hypotheses 3 and 4, it follows that $\Delta \mathbf{F}$
must depend linearly
on $\mathbf{Q},\underline{\underline{\mathbf{\Pi }}}-p_{1}\underline{%
\underline{\mathbf{1}}}.$ On the other hand, again due to
assumption A the vector field $\Delta \mathbf{F}$ must give a
vanishing contribution to the
moments the kinetic equation evaluated with respect to $G(\mathbf{x,}t)=1,%
\mathbf{v,}E\equiv \frac{1}{3}u^{2}.$ Hence, in order that also $\mathbf{F}%
^{\prime }$ depends only on the moments $\left\{ \rho _{o},\mathbf{V,}p_{1},%
\mathbf{Q},\underline{\underline{\mathbf{\Pi }}}\right\} $
(hypothesis 4) necessarily it must result $\Delta
\mathbf{F}(f)\equiv 0$ also for arbitrary non-Maxwellian
distributions $f.$

\section{Fulfillment of the energy equation}

As a further development, let us now impose the additional
requirement that
the inverse kinetic theory yields explicitly also the energy equation (\ref%
{energy equation}). We intend to show that the kinetic equation
fulfilling such a condition can be obtained by a unique
modification of the mean-field force $\mathbf{F\equiv
F}_{0}\mathbf{(x,}t)+\mathbf{F}_{1}\mathbf{(x,}t),$
in particular introducing a suitable new definition of the vector field $%
\mathbf{F}_{1}\mathbf{(x,}t)$. The appropriate new representation
is found to be

\begin{equation*}
\mathbf{F}_{1}\mathbf{(x,}t;f)=\frac{1}{2}\mathbf{u}\frac{\partial \ln p_{1}%
}{\partial t}-\frac{1}{p_{1}}\mathbf{V\cdot }\left\{ \frac{\partial }{%
\partial t}\mathbf{V}+\mathbf{V\cdot \nabla V}+\frac{1}{\rho _{o}}\mathbf{f}%
-\nu \nabla ^{2}\mathbf{V}+\right.
\end{equation*}%
\begin{equation}
\left. +\frac{1}{p_{1}}\mathbf{\nabla \cdot Q}-\frac{1}{p_{1}^{2}}\mathbf{%
\nabla }p\mathbf{\cdot Q}\right\} +\frac{v_{th}^{2}}{2p_{1}}\mathbf{\nabla }%
p\left\{ \frac{u^{2}}{v_{th}^{2}}-\frac{3}{2}\right\} \label{F1
non-maxwellian}
\end{equation}%
As a consequence, the following result holds:\\

\textbf{THEOREM \ 2 -- Inverse kinetic theory for extended INSE}

\emph{Let us require that:}

\emph{1) assumptions 1-3 of Thm.1 are valid;}

\emph{2) the mean-field }$\mathbf{F}$ $\ $\emph{is defined in terms of }$%
\mathbf{F}_{0}$\emph{\ and} $\mathbf{F}_{1}$ \emph{given by Eqs.
(\ref{F0 non-maxwellian case}) and (\ref{F1 non-maxwellian}).}

\emph{Then it follows that:}

\emph{\ A) }$\left\{ \rho ,\mathbf{V,}p\right\} $ \emph{are
classical solutions of extended INSE in }$\Omega \times I$
\emph{if and only if \ the Maxwellian distribution function
}$f_{M}$\emph{\ is a particular solution of the inverse kinetic
equation (\ref{inverse kinetic equation-1});}

\emph{B) provided that the solution }$f(\mathbf{x,}t)$\emph{\ of
the inverse kinetic equation (\ref{inverse kinetic equation-1})
exists in }$\Gamma \times I $ \emph{and results suitably summable
in the velocity space }$U,$ \emph{so that the moment equations}
\emph{of (\ref{inverse kinetic
equation-1}) corresponding to the weight-functions }$G(\mathbf{x,}t)=1,%
\mathbf{v,}E\equiv \frac{1}{3}u^{2}$ \emph{exist}$,$\emph{\ they} \emph{%
coincide necessarily with extended INSE.}

\emph{C) the two representations} \emph{(\ref{F1 non-Maxwellian
case}) and} \emph{(\ref{F1 non-maxwellian}) for }$\mathbf{F}_{1}$
\emph{coincide identically }

\emph{PROOF:}

A) The proof is straightforward. In fact, recalling Thm.1, in \cite%
{Ellero2005}, we notice that Eqs. (\ref{F1 non-maxwellian}) and
(\ref{F1 non-Maxwellian case}) manifestly coincide if and only if
\ the energy equation (\ref{energy equation}) is satisfied
identically, i.e., if the fluid fields are solutions of extended
INSE.

B) The first two moment equations corresponding to $G(\mathbf{x,}t)=1,%
\mathbf{v}$ are manifestly independent of the form of
$\mathbf{F}_{1},$ both in the case of Maxwellian and
non-Maxwellian distributions, i.e., (\ref{F1 non-maxwellian}) and
(\ref{F1 non-Maxwellian case}). Hence, in such a case Thm.3 of
\cite{Ellero2005} applies, i.e., the moment equations yield INSE.
Let us consider, in particular, the third moment equation corresponding to $%
G(\mathbf{x,}t)=\frac{1}{3}u^{2}$,%
\begin{equation}
\frac{\partial }{\partial t}p_{1}+\nabla \cdot \mathbf{Q}+\nabla
\cdot \left[
\mathbf{V}p_{1}\right] -\frac{2}{3}\int d\mathbf{vF(x,}t)\mathbf{u}f+\frac{2%
}{3}\mathbf{\nabla V:\underline{\underline{\Pi }}}=0.
\label{third-moment}
\end{equation}%
Invoking Eqs. (\ref{F0}) and (\ref{F1 non-maxwellian}) for
$\mathbf{F}_{0}$ and $\mathbf{F}_{1},$ the previous equation
reduces to $p_{1}\nabla \cdot \mathbf{V}=0$ if and only if the
energy equation (\ref{energy equation}) is satisfied. Since by
construction $p_{1}>0,$ this yields the isochoricity condition
$\nabla \cdot \mathbf{V}=0.$

C) Finally, since\ thanks to A) $\left\{ \rho
,\mathbf{V,}p\right\} $ are necessarily classical solutions of
INSE, it follows that they fulfill
necessarily also the energy equation (\ref{energy equation}). Hence, (\ref%
{F1 non-Maxwellian case}) and (\ref{F1 non-maxwellian}) coincide
identically in \emph{\ }$\Gamma \times I$.

We conclude that (\ref{F0}) and (\ref{F1 non-maxwellian}) provide
a new form of the inverse kinetic equation applying also to
non-Maxwellian equilibria, which results alternative to that given
earlier in \cite{Ellero2005}. The new form applies necessarily to
classical solutions. Since weak solutions (and hence possibly also
numerical solutions) of INSE may not satisfy exactly the energy
equation, the present inverse kinetic theory based on the new
definition given above [see Eq.(\ref{F1 non-Maxwellian case})] for
the vector field $\mathbf{F(x,}t)$ provides a necessary condition
for the existence of strong solutions of INSE. The result seems
potentially relevant both from the conceptual viewpoint in
mathematical research and for numerical applications.

\section{Conclusions}

In this paper the non-uniqueness of the definition of the inverse
kinetic equation defined by Ellero and Tessarotto (see
\cite{Ellero2005}) has been investigated, proving that the
mean-field force $\mathbf{F}$ characterizing such an equation
depends on an arbitrary real parameter $\alpha .$ To resolve the
indeterminacy, a suitably symmetrization condition has been
introduced for the kinetic energy flux moment equation. As a
consequence, the functional form the mean-field force $\mathbf{F}$
which characterizes the inverse kinetic equation results uniquely
determined. \ Furthermore, we have proven the positivity of the
kinetic distribution function. An open issue remains, however,
whether the inverse kinetic equation (\ref{inverse kinetic
equation-1}) satisfies an H-theorem, i.e., the entropy results a
mononically increasing function of time. Finally, as an additional
development, we have shown that, consistently with the assumption
that the fluid fields are strong solutions of INSE, the mean-field
force can be expressed in such a way to satisfy explicitly also
the energy equation. The result appears significant from the
mathematical viewpoint, the physical interpretation of the theory
and potential applications to the investigation
of complex fluids, such as for example those treated in \cite%
{ellero1,ellero2,ellero3}). In fact, it proves that the inverse
kinetic theory developed in \cite{Ellero2005} can be given an
unique form which applies to classical solutions of INSE.



\section*{Acknowledgments}
Research developed in the framework of MIUR (Ministero
Universit\'a e Ricerca Scientifica, Italy) PRIN Project \textit{\
Fundamentals of kinetic theory and applications to fluid dynamics,
magnetofluiddynamics and quantum mechanics}, partially supported
(P.N.) by CMFD Consortium (Consorzio di Magnetofluidodinamica,
Trieste, Italy).

\end{document}